\documentclass{article}
\usepackage{spconf,amsmath,graphicx,hyperref}
\ninept

\usepackage{color}
\usepackage{epsfig}
\usepackage{graphicx}

\usepackage{booktabs}  %
\usepackage{tabularx}               %
\usepackage{ltablex}

\newcolumntype{C}{>{\centering\arraybackslash}X}
\usepackage{multirow}               %
\usepackage{diagbox}                %
\usepackage{hhline}                 %

\usepackage{extarrows}
\usepackage{makecell}
\usepackage{colortbl}
\usepackage{longtable}
\usepackage[most]{tcolorbox}
\tcbuselibrary{skins} %

\usepackage{amssymb}
\usepackage{amsthm}
\usepackage{amsfonts}

\usepackage{adjustbox}
\usepackage{array}
\usepackage{floatflt}

\usepackage{bm}
\usepackage{nicefrac}
\usepackage{microtype}

\usepackage{changepage}
\usepackage{extramarks}
\usepackage{fancyhdr}
\usepackage{lastpage}
\usepackage{soul}
\usepackage{xspace}

\usepackage{arydshln} %

\usepackage{enumerate}
\usepackage{enumitem}  %

\usepackage{pifont} %

\usepackage{algpseudocode}
\usepackage[ruled,vlined]{algorithm2e} %

\usepackage[symbol]{footmisc}

\usepackage[caption=false]{subfig}

\usepackage{scalefnt}

\usepackage{fontawesome5}

\usepackage{siunitx}

\usepackage{listings}

\newcolumntype{L}[1]{>{\raggedright\let\newline\\\arraybackslash\hspace{0pt}}m{#1}}
\newcolumntype{R}[1]{>{\raggedleft\let\newline\\\arraybackslash\hspace{0pt}}m{#1}}

\newcommand{\ignore}[1]{}

\makeatletter
\DeclareRobustCommand\onedot{\futurelet\@let@token\@onedot}
\def\@onedot{\ifx\@let@token.\else.\null\fi\xspace}

\makeatother

\definecolor{Emerald}{HTML}{50C878}

\definecolor{bestBase}{HTML}{D44478}
\definecolor{secondBase}{HTML}{FFAA33}
\definecolor{thirdBase}{HTML}{FFDD33}

\colorlet{bestLight}{bestBase!20!white}
\colorlet{secondLight}{secondBase!20!white}
\colorlet{thirdLight}{thirdBase!20!white}

\definecolor{MyBlue}{rgb}{0.46, 0.50, 0.61}
\definecolor{MyDarkBlue}{rgb}{0,0.08,0.8}
\definecolor{MyDarkGreen}{RGB}{45,155,45}
\definecolor{MyDarkRed}{rgb}{0.8,0.02,0.02}
\definecolor{MyOrange}{rgb}{1.0, 0.4, 0.2}
\definecolor{MyPurple}{RGB}{111,0,255}
\definecolor{MyRed}{rgb}{0.8,0.0,0.0}
\definecolor{MyGold}{rgb}{0.75,0.6,0.12}
\definecolor{MyDarkgray}{rgb}{0.66, 0.66, 0.66}
\definecolor{MyBrown}{rgb}{0.65, 0.16, 0.16}
\definecolor{MyMutedRose}{rgb}{0.58, 0.29, 0.35}
\definecolor{JiayuanColor}{rgb}{0.60,0.43,0.48}
\definecolor{erranColor}{rgb}{24, 40, 113}

\definecolor{citecolor}{HTML}{696FAD}

\DeclareRobustCommand{\modelname}{\textls[-15]{EVEREST}\xspace}

\newif\ifpropositionfirstitem
\propositionfirstitemtrue

\definecolor{bggray}{HTML}{F5F5F5}
\definecolor{pvdblue}{HTML}{DAE8FC}
\definecolor{RoseQuartzBg}{HTML}{F7CAC9}
\definecolor{RoseQuartz}{HTML}{F5A798}
\definecolor{Serenity}{HTML}{92A8D1}
\definecolor{OrangeRed}{rgb}{1.0, 0.27, 0.0}
\definecolor{RoyalBlue}{cmyk}{1, 0.50, 0, 0}
\definecolor{Turquoise}{HTML}{0F4C81}
\definecolor{mint}{rgb}{0.24, 0.71, 0.54}

\newdimen\abovecrulesep
\newdimen\belowcrulesep
\makeatletter
\patchcmd{\@@@cmidrule}{\aboverulesep}{\abovecrulesep}{}{}
\patchcmd{\@xcmidrule}{\belowrulesep}{\belowcrulesep}{}{}
\makeatother

\definecolor{mybluetitle}{HTML}{4B527E} %

\definecolor{mygreen}{RGB}{0,150,0}
\definecolor{boxbackground}{HTML}{F0F7FF}  %
\definecolor{boxborder}{HTML}{D0D9E5}      %
\definecolor{accentblue}{HTML}{4A86E8}     %
\definecolor{lightblue}{HTML}{EEF3FF}  %
\definecolor{bordergray}{HTML}{CCCCCC}  %
\definecolor{headerblue}{HTML}{2C5AA0}  %

\definecolor{lavenderframe}{HTML}{E6E6FA}  %
\definecolor{lighterlav}{HTML}{F5F5FF}  %
\definecolor{codegray}{rgb}{0.5,0.5,0.5}  %
\definecolor{codepurple}{HTML}{483D8B}  %
\definecolor{backcolour}{HTML}{F5F5FF}  %
\lstdefinestyle{mystyle}{
    backgroundcolor=\color{backcolour},
    commentstyle=\color{headerblue},
    keywordstyle=\color{codepurple},
    numberstyle=\tiny\color{codegray},
    stringstyle=\color{codepurple},
    basicstyle=\ttfamily\scriptsize,
    breakatwhitespace=false,
    breaklines=true,
    captionpos=b,
    keepspaces=true,
    frame=none,
    numbersep=5pt,
    showspaces=false,
    showstringspaces=false,
    showtabs=false,
    tabsize=2
}

\definecolor{jsonkey}{RGB}{44, 130, 201}     %
\definecolor{jsonstring}{RGB}{255, 140, 0}   %
\definecolor{jsonnumber}{RGB}{34, 139, 34}   %

\lstdefinelanguage{json}{
    basicstyle=\ttfamily\small,
    numbers=left,
    numberstyle=\tiny\color{gray},
    stepnumber=1,
    numbersep=5pt,
    showstringspaces=false,
    breaklines=true,
    frame=none,
    backgroundcolor=\color{gray!5},
    literate=
     *{:}{{{\color{jsonkey}:}}}{1}
      {,}{{{\color{jsonkey},}}}{1}
      {"}{{{\color{jsonstring}"}}}{1}
      {[}{{{\color{jsonkey}[}}}{1}
      {]}{{{\color{jsonkey}]}}}{1}
      {0}{{{\color{jsonnumber}0}}}{1}
      {1}{{{\color{jsonnumber}1}}}{1}
      {2}{{{\color{jsonnumber}2}}}{1}
      {3}{{{\color{jsonnumber}3}}}{1}
      {4}{{{\color{jsonnumber}4}}}{1}
      {5}{{{\color{jsonnumber}5}}}{1}
      {6}{{{\color{jsonnumber}6}}}{1}
      {7}{{{\color{jsonnumber}7}}}{1}
      {8}{{{\color{jsonnumber}8}}}{1}
      {9}{{{\color{jsonnumber}9}}}{1}
}

\newtcblisting{jsonbox}{
  listing engine=listings,
  colback=gray!3!white,
  colframe=gray!75!black,
  boxrule=0.4mm,
  arc=2mm,
  outer arc=2mm,
  breakable,
  enhanced,
  listing only,
  listing options={language=json}
}

\newtcolorbox{promptbox}[2][]{ %
    enhanced,
    breakable,
    boxsep=5pt,
    left=9pt,
    right=7pt,
    top=5pt,
    bottom=5pt,
    colback=boxbackground,
    colframe=boxborder,
    boxrule=0.5pt,
    arc=4pt,
    frame hidden, %
    borderline west={3pt}{0pt}{accentblue},
    shadow={0.5pt}{0.5pt}{1.5pt}{black!10},
    fontupper=\normalsize,
    title=#2, %
    colbacktitle=accentblue, %
    coltitle=white,         %
    fonttitle={\fontsize{9}{11}\selectfont\bfseries}, %
    attach boxed title to top left={yshift=-2.5mm, xshift=3.2mm},
    boxed title style={
        enhanced,
        left=3pt,
        right=3pt,
        top=1pt,    %
        bottom=1pt, %
        boxsep=2pt,
        arc=3pt,
        boxrule=0pt,
        colback=accentblue,
    },
    #1 %
}

\newtcolorbox{notitlepromptbox}[1][]{
    enhanced,
    breakable,
    boxsep=5pt,          %
    left=9pt,            %
    right=7pt,           %
    top=5pt,             %
    bottom=5pt,          %
    colback=boxbackground,
    colframe=boxborder,
    boxrule=0.5pt,
    arc=4pt,             %
    frame hidden,
    borderline west={3pt}{0pt}{accentblue},  %
    shadow={0.5pt}{0.5pt}{1.5pt}{black!10},  %
    notitle,
    fontupper=\normalsize,    %
    #1
}

\newtcolorbox{onebox}[2][]{
    enhanced, 
    center title,
    left*=0pt, right*=0pt,
    boxsep=2pt, left=5pt, right=5pt,
    skin first=enhanced,
    skin middle=enhanced,
    skin last=enhanced,
    colframe = mybluetitle!90,
  colback  = mybluetitle!10,
    fonttitle=\bfseries\rmfamily\fontfamily{phv}\selectfont,
    title={\footnotesize\strut{#2}  \refstepcounter{subsubsection} \addcontentsline{toc}{subsubsection}{\string\numberline{\thesubsubsection}#2}
    },
    #1
    }

\colorlet{osfirst}{teal!50}
\colorlet{ossecond}{teal!30}
\colorlet{osthird}{teal!10}
\colorlet{lavenderfirst}{violet!50}
\colorlet{lavendersecond}{violet!30}
\colorlet{lavenderthird}{violet!10}

\theoremstyle{plain}

\theoremstyle{definition}

\theoremstyle{remark}

\definecolor{lightgray}{rgb}{0.88, 0.92, 0.98}
\definecolor{defblue}{rgb}{0.1843, 0.3333, 0.6}
\definecolor{defred}{rgb}{0.88, 0.2510, 0.3294}

\definecolor{green1}{rgb}{ 0.910,  0.953,  0.855}
\definecolor{green2}{rgb}{0.82,  0.902,  0.710}
\definecolor{green3}{rgb}{0.713,  0.903,  0.648}
\definecolor{green4}{rgb}{ 0.725,  0.855,  0.561}

\definecolor{defyellow}{rgb}{1,  0.983,  0.717}
\definecolor{defyellowtext}{rgb}{1,  0.851,  0.438}

\title{\modelname: Endogenous Vision-Language Reinforcement Reasoning Exploration for Urban Socio-Semantic Segmentation}
\name{Qixiu Li$^{1,\dagger}$
        \! Zhongzhi He$^{2,\dagger}$
        \! Xiang Zhu$^{1,*}$
        \! Xiaoyong Li$^{1,*}$\thanks{$\dagger$ Equal Contribution. *Corresponding Author.}
        \! Jiarun Lin$^{1}$
        \! Weifeng Xu$^{1}$}

\address{$^{1}$National University of Defense Technology \\
         \! $^{2}$Changchun University of Science and Technology}

\begin{document}
%
\maketitle

\begin{abstract}
Urban socio-semantic segmentation leverages digital and satellite imagery to provide critical spatial semantic information for downstream applications such as urban resource allocation. Although existing methods achieve high segmentation accuracy, they still suffer from inaccurate delineation of target boundaries. The underlying issue is that current models primarily rely on passively aggregated global cross-modal cues, lacking active exploration of the environment. To address this limitation, we propose the \modelname model, which adopts an egocentric exploration strategy that enables the model to actively investigate boundary cues and perform self-correction. In addition, we formulate discrete natural-language prompts as pseudocode to regularize the execution logic. Reinforcement learning is further employed to implement this irreducible process and elicit the model’s structured reasoning capability. Our \modelname achieves optimal performance on all metrics in the real world urban socio-semantic dataset, demonstrating the superiority of our model. Code is available at~\href{https://github.com/TechCloud-x/EVEREST}{https://github.com/TechCloud-x/EVEREST}.
\end{abstract}
\vspace{-5pt}

\vspace{-1.2em}
\section{Introduction}\label{sec:intro}
\vspace{-0.9em}
Urban socio-semantic segmentation aims to recover the pixel-level extent of socially defined urban entities from multimodal geospatial observations~\cite{reades2025city}. Unlike conventional land-cover objects, urban regions contain not only buildings, roads, and vegetation~\cite{nilforoshan2023human}, but also schools, hospitals, parks, residential and commercial districts whose identities are determined by social functions and place names. Accurate boundaries of these entities are essential for urban resource allocation, navigation services, Area of Interest management, and proximity-based planning such as the 15-minute city~\cite{bruno2024universal,shi2025multimodal}. Therefore, this task pushes remote sensing interpretation from recognizing physical surfaces toward understanding functional urban spaces.

\begin{figure}[t!]
  \begin{center}
\includegraphics[width=.86\linewidth]{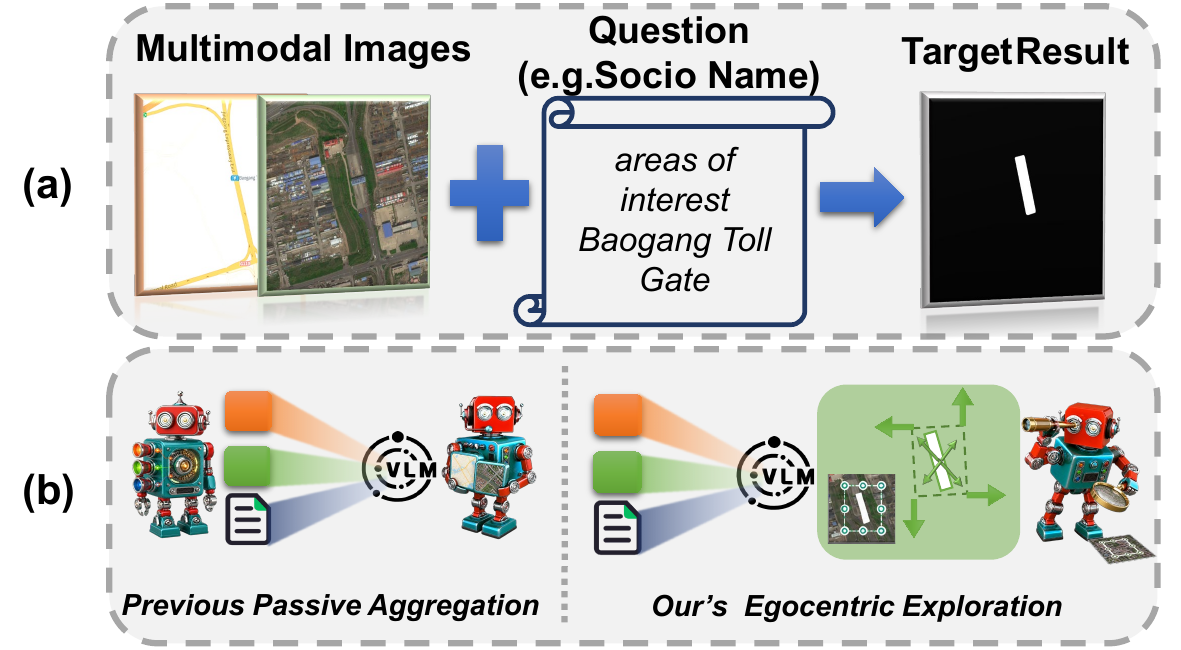}
  \end{center}
    \vspace{-22pt}
  \caption{(a) An example of the urban socio-semantic task. (b) The difference between previous work and \modelname's main idea.}
  \vspace{-10pt}
  \label{fig:teaser}
  \vspace{-10pt}
\end{figure}

Existing remote sensing segmentation methods have achieved remarkable progress on visually defined categories. CNN-based models such as UNet~\cite{ronneberger2015u} and Transformer-based~\cite{vaswani2017attention} models such as SegFormer~\cite{xie2021segformer} learn dense visual representations from image appearance, while recent open-vocabulary approaches such as SegEarth-OV~\cite{li2025segearthov} further relax the closed-set category assumption. However, these methods are still largely appearance-driven. They are well suited to entities with stable spectral, textural, or geometric cues, but struggle when the target is defined by a name, function, or socioeconomic role (Fig. \ref{fig:teaser} (a)) rather than by visible shape alone.

Digital maps provide a natural source of social semantic signals. Map tiles render place names, road structures, points of interest, and cartographic symbols into a visual layer that is spatially aligned with satellite imagery. Recent vision-language models (VLMs)~\cite{oelsmann2024regional,he2026rethinking} and promptable segmentation models make it possible to jointly reason over maps, satellite images, and textual instructions, and then convert visual prompts into masks with SAM-like segmenters~\cite{ravi2025sam,chen2025rsrefseg,yao2026remotereasoner}. SocioReasoner~\cite{wang2026urban} first demonstrates this paradigm for urban socio-semantic segmentation. Nevertheless, most existing pipelines~\cite{li2025segearth,huang2025samr1,you2025segr1,liu2025visionreasoner} still follow a passive ``observe-then-predict'' pattern: the model aggregates global cross-modal cues and emits boxes or points, but rarely performs explicit boundary-oriented exploration.

This passive formulation leaves a concrete gap (Fig. \ref{fig:teaser} (b)). Although map-satellite reasoning helps localize socially meaningful regions, current models~\cite{chen2025cost,li2026frequency} lack a mechanism for actively inspecting candidate boundaries and confirming each target instance. As a result, initial localization errors can propagate into the final mask, especially in dense urban scenes where map labels are crowded, satellite textures are ambiguous, or multiple instances share the same social category. In addition, natural-language prompts are often underspecified as execution policies. Their loose structure makes it difficult to reliably transform free-form reasoning into executable visual primitives such as boxes, points, and keep-or-adjust decisions.

To address these limitations, we propose \textbf{\name}, an \textbf{\underline{E}}ndogenous \textbf{\underline{V}}ision-languag\textbf{\underline{E}} \textbf{\underline{R}}einforcem\textbf{\underline{E}}nt rea\textbf{\underline{S}}oning explora\textbf{\underline{T}}ion method for urban socio-semantic segmentation. \ding{182} \modelname introduces an egocentric exploration strategy that encourages the VLM to enumerate candidate instances, inspect rendered feedback, and refine target boundaries through self-correction. \ding{183} To regularize this process, we formulate natural-language instructions as structured pseudocode, so that the model's reasoning can be converted into stable visual primitives for SAM-based segmentation. This design treats urban segmentation as an interactive multimodal signal processing problem, where map semantics, satellite appearance, and executable prompts are jointly optimized. \ding{184} Since the resulting VLM-SAM workflow is non-differentiable, we optimize it with reinforcement learning, using rewards that align structured reasoning, localization, instance consistency, and final mask quality. \ding{185} Experiments on the real-world SocioSeg benchmark show that \modelname outperforms competitive remote sensing and reasoning-segmentation baselines.

\begin{figure*}[ht]
\vspace{-10pt}
  \begin{center}
  \includegraphics[width=0.9\linewidth]{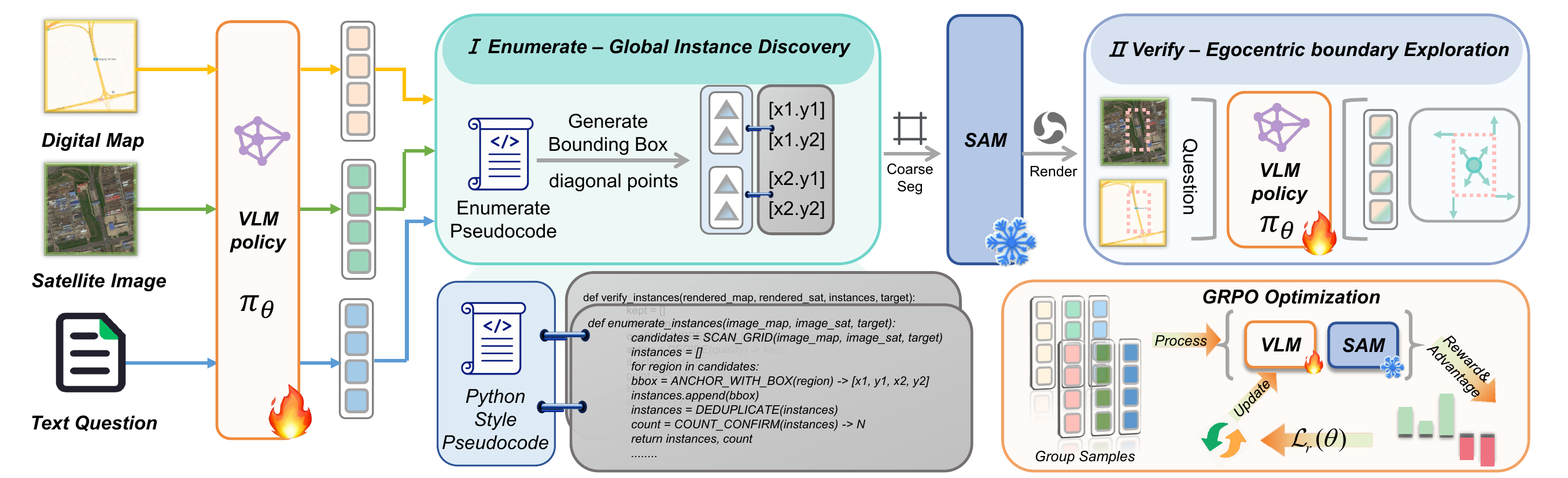}
  \end{center}
  \vspace{-20pt}
  \caption{\small The framework of our proposed \modelname. We design enumerate and verify two stages respectively.}
  \vspace{-15pt}
  \label{fig:main}
\end{figure*}

\section{Methodology}\label{sec:method}

\noindent We propose \modelname, an endogenous vision-language reinforcement reasoning exploration framework for urban socio-semantic segmentation. In Fig.\ref{fig:main}, given a digital map, a satellite image, and a textual target, \modelname converts cross-modal reasoning into executable visual primitives. The framework contains two coupled stages. The \textit{Enumerate} stage scans the whole scene and assigns each candidate social entity a stable instance identity with a bounding box. The \textit{Verify} stage observes rendered feedback, performs egocentric boundary exploration around each candidate, and decides whether to keep, adjust, or drop the instance. The final visual primitives are executed by a frozen SAM2 segmenter~\cite{ravi2025sam}. Because parsing, rendering, and SAM2 execution are non-differentiable, the VLM policy is optimized with GRPO-style reinforcement learning~\cite{guo2025deepseek,song2026balancing,han2026long}.

\subsection{Problem Formulation}
Let $\mathbf{I}_m,\mathbf{I}_s\in\mathbb{R}^{H\times W\times 3}$ denote a rendered digital map and its spatially aligned satellite image, and let $\mathbf{q}$ be a text instruction describing a social semantic target, e.g., a name, a class, or a functional zone. The goal is to predict a binary mask $\hat{\mathbf{M}}\in\{0,1\}^{H\times W}$ that matches the ground-truth mask $\mathbf{M}^{\star}$. Instead of directly regressing the mask, \modelname learns a VLM policy $\pi_{\theta}$ that emits a structured textual output $\mathbf{y}$ containing a reasoning program and a JSON answer. A parser $\mathcal{P}$ maps the answer into visual primitives, and a frozen segmenter $\mathcal{S}$ converts them into masks. Formally, these can be expressed as:
\begin{equation}
\mathbf{z}=\mathcal{P}(\mathbf{y}),\quad
\hat{\mathbf{M}}=\mathcal{S}(\mathbf{I}_s,\mathbf{z}),\quad
\mathbf{y}\sim\pi_{\theta}(\cdot\mid \mathbf{I}_m,\mathbf{I}_s,\mathbf{q}).
\end{equation}
The optimization objective is therefore to learn a policy that maximizes a task-level quality function $Q(\hat{\mathbf{M}},\mathbf{M}^{\star})$ while preserving valid executable structure, which can be expressed as:
\begin{equation}
\theta^{\star}=\arg\max_{\theta}\;\mathbb{E}_{\mathbf{y}\sim\pi_{\theta}}
\left[Q\left(\mathcal{S}(\mathbf{I}_s,\mathcal{P}(\mathbf{y})),\mathbf{M}^{\star}\right)\right].
\end{equation}

\subsection{Endogenous Visual Primitive Exploration}
\noindent\textbf{Pseudocode-Guided Instance Enumeration.}
The Enumerate stage transforms global map-satellite perception into instance-level visual primitives. Given $\mathbf{x}_E=(\mathbf{I}_m,\mathbf{I}_s,\mathbf{q})$, the policy emits an executable pseudocode trace constrained by four operators: scan\_grid, anchor\_with\_box, deduplicate, and 
count\_confirm. The answer is parsed into a visual primitive state, which is mapping as:
\begin{equation}
\mathbf{Z}_E=
\left\{\mathbf{q},\hat{N},\left(i,\mathbf{b}_i,\mathbf{c}_i,e_i,\rho_i\right)_{i=1}^{\hat{N}}\right\},
\end{equation}
where $i$ is a stable instance index, $\mathbf{b}_i=[x_i^1,y_i^1,x_i^2,y_i^2]$ is a pixel-coordinate bounding box, $\mathbf{c}_i=[(x_i^1+x_i^2)/2,(y_i^1+y_i^2)/2]$ is the center point, $e_i$ records the evidence source, and $\rho_i$ denotes confidence. The valid coordinate domain is formulated as:
\begin{equation}
\begin{aligned}
\Omega &= [0,W-1]\times[0,H-1],\\
\mathcal{B}_{\Omega} &=
\{[x^1,y^1,x^2,y^2]\mid
0\leq x^1<x^2<W,\\
&\hspace{7.4em}0\leq y^1<y^2<H\}.
\end{aligned}
\end{equation}
Each $\mathbf{b}_i$ is executed as a box prompt by SAM2, and the coarse mask is the union of all instance masks, which can be expressed as:
\begin{equation}
\mathbf{M}_c=\bigvee_{i=1}^{\hat{N}}\mathcal{S}(\mathbf{I}_s,\mathbf{b}_i).
\end{equation}
This instance-indexed representation prevents the model from treating multiple targets as an anonymous list of boxes, making later verification explicitly referential.

\noindent\textbf{Rendered Feedback and Egocentric Verification.}
The coarse prediction is rendered back onto both modalities through a deterministic renderer $\mathcal{D}$. This process is mathematically formulated as:
\begin{equation}
(\widetilde{\mathbf{I}}_m,\widetilde{\mathbf{I}}_s)
=\mathcal{D}(\mathbf{I}_m,\mathbf{I}_s,\mathbf{M}_c,\{(i,\mathbf{b}_i,\mathbf{c}_i)\}_{i=1}^{\hat{N}}).
\end{equation}
The Verify stage then inspects every indexed primitive rather than generating a new unordered prediction. Its pseudocode is constrained by inspect, decide, place\_point, and verify\_count. For egocentric exploration, we define the boundary anchor set of $\mathbf{b}_i$ as:
\begin{equation}
\begin{aligned}
\bar{x}_i &= \frac{x_i^1+x_i^2}{2},\qquad
\bar{y}_i = \frac{y_i^1+y_i^2}{2},\\
\mathcal{C}_i &= \{(x_i^1,y_i^1),(x_i^2,y_i^1),
(x_i^2,y_i^2),(x_i^1,y_i^2)\},\\
\mathcal{E}_i &= \{(\bar{x}_i,y_i^1),(x_i^2,\bar{y}_i),
(\bar{x}_i,y_i^2),(x_i^1,\bar{y}_i)\},\\
\mathcal{A}(\mathbf{b}_i) &= \mathcal{C}_i\cup\mathcal{E}_i .
\end{aligned}
\end{equation}
The counter-clockwise angular exploration field is specified as an ordered sequence. This operation can be formulated as:
\begin{equation}
\begin{aligned}
\mathbf{R}(\alpha)&=
\begin{bmatrix}
\cos\alpha&-\sin\alpha\\
\sin\alpha& \cos\alpha
\end{bmatrix},\\
\Theta_{\perp}&=\left(0,\frac{\pi}{2},\pi,\frac{3\pi}{2}\right),\\
\Theta_{\diamond}&=\left(\frac{\pi}{4},\frac{3\pi}{4},
\frac{5\pi}{4},\frac{7\pi}{4}\right).
\end{aligned}
\end{equation}
\begin{equation}
\begin{gathered}
\Theta = \Theta_{\perp}\oplus\Theta_{\diamond},\ 
\mathbf{u}(\alpha)=\mathbf{R}(\alpha)
\begin{bmatrix}1\\0\end{bmatrix},\ 
L_i=\|\mathbf{b}_i^{\mathrm{br}}-\mathbf{b}_i^{\mathrm{tl}}\|_2,\\[0.5em]
\Gamma_i=\Big\{
\operatorname{clip}_{\Omega}
\big(\mathbf{a}+\lambda\mathbf{u}(\alpha)\big)
\;\Big|\;\\
\mathbf{a}\in\mathcal{A}(\mathbf{b}_i),\;
\alpha\in\Theta,\;
0\leq\lambda\leq L_i
\Big\}.
\end{gathered}
\end{equation}
Given a learned VLM scoring function $\psi_{\theta}$ over rendered evidence, the positive exploration points are abstractly written as:
\begin{equation}
\mathcal{P}_i^{+}=\operatorname{TopK}_{\mathbf{p}\in\Gamma_i}\;
\psi_{\theta}(\widetilde{\mathbf{I}}_m,\widetilde{\mathbf{I}}_s,\mathbf{q},i,\mathbf{b}_i,\mathbf{p}).
\end{equation}
The Verify output is computed mathematically as:
\begin{equation}
\mathbf{Z}_V=\left\{(i,a_i,\widetilde{\mathbf{b}}_i,\mathcal{P}_i^{+})\right\}_{i=1}^{\hat{N}},\quad
a_i\in\{\mathtt{keep},\mathtt{adjust},\mathtt{drop}\}.
\end{equation}

\subsection{Mask Generation with Instance-Aware Prompt Merging}
The final prompt set is obtained by merging Enumerate primitives and Verify decisions by instance identity. For index $i$, the box is:
\begin{equation}
\bar{\mathbf{b}}_i=
\begin{cases}
\mathbf{b}_i, & a_i=\mathtt{keep},\\
\widetilde{\mathbf{b}}_i, & a_i=\mathtt{adjust}\;\land\;\widetilde{\mathbf{b}}_i\in\mathcal{B}_{\Omega},\\
\varnothing, & a_i=\mathtt{drop}.
\end{cases}
\end{equation}
If Verify omits an Enumerate instance, \modelname conservatively keeps the original box. The final executable primitive set is:
\begin{equation}
\mathcal{G}=\left\{(\bar{\mathbf{b}}_i,\mathcal{P}_i^{+})\mid \bar{\mathbf{b}}_i\neq\varnothing\right\},
\end{equation}
which is sent to SAM2 as box-plus-positive-point prompts. The predicted mask is the instance-level union. The process is as follow:
\begin{equation}
\hat{\mathbf{M}}=\bigvee_{(\bar{\mathbf{b}}_i,\mathcal{P}_i^{+})\in\mathcal{G}}
\mathcal{S}(\mathbf{I}_s,\bar{\mathbf{b}}_i,\mathcal{P}_i^{+}).
\end{equation}
This merging rule preserves stable references across stages, suppresses false positives through drop, and allows boundary correction through adjust and positive-point exploration.

\subsection{Reinforcement Learning Optimization}
The EVEREST pipeline contains discrete text generation, JSON parsing, rendering, visual prompt construction, and SAM2~\cite{ravi2025sam} execution. These operations prevent direct back-propagation from the final mask to the VLM. We therefore optimize the shared policy $\pi_{\theta}$ with group-relative policy optimization. For each stage $r\in\{E,V\}$ and each input $\mathbf{x}_r$, a group of $G$ responses is sampled from the behavior policy. The environment executes the parsed visual primitives and returns a scalar reward $R_r^g$, which summarizes structural validity, localization quality, instance consistency, point usefulness, and final mask accuracy. We keep the reward design as an external evaluator and use it only to construct relative advantages for policy learning. The group statistics and normalized advantage are computed by as follow:
\begin{equation}
\mu_r = \frac{1}{G}\sum_{g=1}^{G}R_r^g, \ 
\sigma_r^2 = \frac{1}{G}\sum_{g=1}^{G}(R_r^g-\mu_r)^2, \ 
A_r^g = \frac{R_r^g-\mu_r}{\sigma_r+\epsilon}.
\end{equation}
For token $t$ in output $\mathbf{y}_r^g$, the policy ratio is defined as:
\begin{equation}
\omega_{r,t}^{g}(\theta)=
\frac{\pi_{\theta}(y_{r,t}^{g}\mid \mathbf{x}_r,y_{r,<t}^{g})}
{\pi_{\theta_{\mathrm{old}}}(y_{r,t}^{g}\mid \mathbf{x}_r,y_{r,<t}^{g})}.
\end{equation}
The clipped token-level surrogate is mathematically formulated as:
\begin{equation}
\begin{aligned}
\ell_{r,t}^{g}(\theta)=
\min\Big(&\omega_{r,t}^{g}(\theta)A_r^g,\\
&\operatorname{clip}(\omega_{r,t}^{g}(\theta),1-\eta,1+\eta)A_r^g\Big).
\end{aligned}
\end{equation}
Thus, the GRPO loss for stage $r$ is formulated as:
\begin{equation}
\begin{aligned}
\mathcal{L}_r(\theta)=
-&\frac{1}{G}\sum_{g=1}^{G}
\frac{1}{|\mathbf{y}_r^g|}\sum_t \ell_{r,t}^{g}(\theta)\\
&+\beta\,\mathrm{KL}\left(
\pi_{\theta}(\cdot\mid\mathbf{x}_r)\,\|\,
\pi_{\mathrm{ref}}(\cdot\mid\mathbf{x}_r)
\right).
\end{aligned}
\end{equation}
The overall objective jointly optimizes Enumerate and Verify with shared VLM parameters formally can be expressed as:
\begin{equation}
\theta^{\star}=\arg\min_{\theta}
\left[\mathcal{L}_E(\theta)+\mathcal{L}_V(\theta)\right].
\end{equation}
This loss couples the two stages without differentiating through the external segmenter: Enumerate learns to produce stable and complete primitive anchors, while Verify learns to refine them into executable prompts that improve final socio-semantic masks.
\begin{table*}[th]
  \centering
  \caption{Comparison with state-of-the-art methods on SocioSeg test set. Top three results are highlighted as \colorbox{bestLight}{\textbf{best}}, \colorbox{secondLight}{second}, and \colorbox{thirdLight}{third}.}
  \label{tab1:results}
  \tiny
  \resizebox{.9\textwidth}{!}{%
  \begin{tabular}{l c c c c c c c c c c}
    \toprule
    \multirow{2}{*}{Method}
    & \multirow{2}{*}{Proc.\ \& Year}
    & \multicolumn{2}{c}{Socio-name}
    & \multicolumn{2}{c}{Socio-class}
    & \multicolumn{2}{c}{Socio-function}
    & \multicolumn{2}{c}{All dataset}
    & \multirow{2}{*}{Avg. \ Rank} \\

    \cmidrule(lr){3-4}
    \cmidrule(lr){5-6}
    \cmidrule(lr){7-8}
    \cmidrule(lr){9-10}

    & & cIoU $\uparrow$ & F1 $\uparrow$
    & cIoU $\uparrow$ & F1 $\uparrow$
    & cIoU $\uparrow$ & F1 $\uparrow$
    & cIoU $\uparrow$ & F1 $\uparrow$
    & \\

    \midrule

    UNet~\cite{ronneberger2015u}
    & MICCAI 2015
    & 10.9
    & 8.0
    & 12.6
    & 11.2
    & 11.1
    & 10.4
    & 11.7
    & 10.0
    & \cellcolor{gray!30}10 \\

    Segformer~\cite{xie2021segformer}
    & NeurIPS 2021
    & 22.0
    & 18.1
    & 22.4
    & 19.5
    & 21.4
    & 17.9
    & 22.1
    & 18.7
    & \cellcolor{gray!30}9 \\

    \midrule

    VisionReasoner~\cite{liu2025visionreasoner}
    & ICLR 2026
    & \cellcolor{thirdLight}48.5
    & \cellcolor{thirdLight}58.4
    & \cellcolor{thirdLight}44.4
    & \cellcolor{thirdLight}55.5
    & 36.3
    & 45.0
    & \cellcolor{thirdLight}44.0
    & \cellcolor{thirdLight}54.3
    & \cellcolor{gray!30}3 \\

    Seg-R1~\cite{you2025segr1}
    & NeurIPS 2025
    & 46.0
    & 50.4
    & 40.4
    & 45.2
    & 34.5
    & 36.5
    & 41.0
    & 45.2
    & \cellcolor{gray!30}6 \\

    SAM-R1~\cite{huang2025samr1}
    & NeurIPS 2025
    & 25.6
    & 37.2
    & 22.3
    & 32.1
    & 17.7
    & 25.2
    & 22.5
    & 32.4
    & \cellcolor{gray!30}8 \\

    \midrule

    SegEarth-OV~\cite{li2025segearthov}
    & CVPR 2025
    & 3.3
    & 0.0
    & 3.8
    & 0.0
    & 4.2
    & 0.0
    & 3.7
    & 0.0
    & \cellcolor{gray!30}11 \\

    RSRefSeg~\cite{chen2025rsrefseg}
    & IGARSS 2025
    & 27.1
    & 30.9
    & 30.7
    & 35.3
    & 28.7
    & 30.8
    & 29.0
    & 32.8
    & \cellcolor{gray!30}7 \\

    SegEarth-R1~\cite{li2025segearth}
    & arXiv 2025
    & 36.9
    & 46.9
    & 38.9
    & 50.0
    & \cellcolor{thirdLight}39.5
    & \cellcolor{thirdLight}47.4
    & 38.3
    & 48.4
    & \cellcolor{gray!30}5 \\

    RemoteReasoner~\cite{yao2026remotereasoner}
    & AAAI 2026
    & 46.6
    & 56.1
    & 42.9
    & 53.9
    & 38.0
    & 47.2
    & 43.2
    & 53.3
    & \cellcolor{gray!30}4 \\

    \midrule

    SocioReasoner~\cite{wang2026urban}
    & ICLR 2026
    & \cellcolor{secondLight}52.6
    & \cellcolor{bestLight}\textbf{64.6}
    & \cellcolor{secondLight}47.6
    & \cellcolor{secondLight}60.1
    & \cellcolor{secondLight}40.6
    & \cellcolor{secondLight}50.3
    & \cellcolor{secondLight}47.9
    & \cellcolor{secondLight}59.7
    & \cellcolor{gray!30}2 \\
    
    \textbf{\modelname (Ours)}
    & --
    & \cellcolor{bestLight}\textbf{53.0}
    & \cellcolor{secondLight}64.3
    & \cellcolor{bestLight}\textbf{50.5}
    & \cellcolor{bestLight}\textbf{61.5}
    & \cellcolor{bestLight}\textbf{44.4}
    & \cellcolor{bestLight}\textbf{54.9}
    & \cellcolor{bestLight}\textbf{50.4}
    & \cellcolor{bestLight}\textbf{61.4}
    & \cellcolor{gray!30}1 \\

    \bottomrule
  \end{tabular}%
  }
  \vspace{-10pt}
\end{table*}
\section{Experiments}\label{sec:experiments}
\subsection{Experimental Settings}
\noindent\textbf{Datasets}. 
We conduct experiments on SocioSeg~\cite{wang2026urban}, a real-world benchmark for urban socio-semantic segmentation. Each sample contains a satellite image, a spatially aligned digital map, and a pixel-level mask of the target social entity. SocioSeg organizes urban entities into three hierarchical task levels: \textit{Socio-name}, which targets a specific named Area of Interest (AOI); \textit{Socio-class}, which describes a fine-grained category such as school or hospital; and \textit{Socio-function}, which refers to more abstract functional regions such as educational or commercial areas. The dataset contains over 13,000 samples, covering more than 5,000 entity names, 90 classes, and 10 functions. Following the Previous work~\cite{wang2026urban}, we evaluate cIoU and F1 for each task group and report their overall averages.

\noindent\textbf{Baselines and Evaluation Metrics}.
We compare \modelname with three groups of representative methods for a comprehensive evaluation. The first group includes conventional semantic segmentation models, i.e., UNet~\cite{ronneberger2015u} and SegFormer~\cite{xie2021segformer}, which mainly rely on visual appearance. The second group contains recent reasoning segmentation models, including VisionReasoner~\cite{liu2025visionreasoner}, Seg-R1~\cite{you2025segr1}, and SAM-R1~\cite{huang2025samr1}. The third group covers remote sensing segmentation and geospatial reasoning methods, including SegEarth-OV~\cite{li2025segearthov}, RSRefSeg~\cite{chen2025rsrefseg}, SegEarth-R1~\cite{li2025segearth}, and RemoteReasoner~\cite{yao2026remotereasoner}. We also include SocioReasoner~\cite{wang2026urban} as the strongest task-specific baseline. For evaluation, cIoU measures region-level mask overlap, while F1 evaluates pixel-level precision-recall balance. Avg. Rank summarizes the relative ranking across all reported metrics.

\noindent\textbf{Implementation Details}. 
To assess the effectiveness of each method, we initially identified the best hyperparameters with a single seed applied to the validation set. Subsequently, each method was trained with the selected hyperparameters across five different random seeds. For \modelname, we use Qwen2.5-VL-3B~\cite{bai2025qwen2} as the vision-language backbone and SAM2~\cite{ravi2025sam} as the frozen promptable segmenter. The reinforcement learning configuration uses a rollout batch size of 128 and a group size of 8. We optimize the VLM policy with AdamW~\cite{loshchilov2019decoupled} using a learning rate of $1\times10^{-6}$. The clipping coefficient is set to $0.5$, and the KL regularization weight is set to $0.005$. We conduct all experiments on four NVIDIA A800 GPUs with 80GB memory.

\subsection{Performance Comparison}
Table~\ref{tab1:results} reports the comparison results on the SocioSeg test set. \textit{1)} Overall, \modelname achieves the best Avg. Rank and obtains the highest overall cIoU and F1, reaching 50.4 and 61.4. Compared with the previous task-specific baseline SocioReasoner, \modelname improves the overall cIoU from 47.9 to 50.4 and the F1 from 59.7 to 61.4. This demonstrates that endogenous exploration and pseudocode-guided visual primitive reasoning provide more reliable target localization and mask generation than passive map-satellite reasoning. \textit{2)} The improvements are especially clear on the more abstract hierarchical tasks. On Socio-class, \modelname achieves 50.5 cIoU and 61.5 F1, surpassing SocioReasoner by 2.9 and 1.4 points. On Socio-function, the gains are larger: \modelname improves cIoU from 40.6 to 44.4 and F1 from 50.3 to 54.9. These results indicate that active verification is particularly useful when targets are defined by functional semantics rather than distinctive visual appearance. Although SocioReasoner obtains a slightly higher F1 on Socio-name, \modelname still achieves the best cIoU on this task, suggesting more accurate boundary delineation for named AOIs. \textit{3)} Compared with conventional segmentation models, \modelname shows a substantial advantage because social entities cannot be reliably inferred from satellite appearance alone. It also outperforms open-vocabulary and referring remote sensing methods, whose predictions remain limited when map semantics and boundary reasoning are required. Among VLM-based reasoning models, VisionReasoner and RemoteReasoner perform competitively, but they still follow a largely observe-then-predict paradigm. \modelname explicitly enumerates candidate instances, renders intermediate evidence, and verifies each target through egocentric exploration, leading to more robust segmentation across all SocioSeg task levels.

\begin{table}[t]
\vspace{-5.5pt}
  \centering
  \tiny
  \caption{Ablation studies of point number and multi-stage design.}
  \label{tab:ablation}
  \resizebox{.8\linewidth}{!}{%
    \begin{tabular}{l c c c c}
      \toprule
      \multirow{2}{*}{Method}
      & \multicolumn{4}{c}{All dataset} \\
      \cmidrule(lr){2-5}
      & cIoU $\uparrow$ & $\Delta$ & F1 $\uparrow$ & $\Delta$ \\
      \midrule

      1 point refinement
      & \cellcolor{thirdLight} 49.5
      & -0.9
      & \cellcolor{thirdLight} 60.6
      & -0.8 \\

      3 points refinement
      & \cellcolor{bestLight} \textbf{50.6}
      & +0.2
      & \cellcolor{secondLight} 61.2
      & -0.2 \\

      4 points refinement
      & 47.7
      & -2.7
      & 59.6
      & -1.8 \\

      \midrule

      w/o Enumeration
      & 43.5
      & -6.9
      & 53.4
      & -8 \\

      w/o Verification
      & 46.3
      & -4.1
      & 58.8
      & -2.6 \\

      \midrule

      \textbf{\modelname (Ours)}
      & \cellcolor{secondLight} 50.4
      & 0.00
      & \cellcolor{bestLight} \textbf{61.4}
      & 0.00 \\

      \bottomrule
    \end{tabular}%
  }
  \vspace{-10pt}
\end{table}

\subsection{Ablation Study}
\noindent We conduct ablation studies from two perspectives: the number of positive refinement points and the necessity of the two-stage exploration design. As shown in Table~\ref{tab:ablation}, the full \modelname obtains the best F1 score of 61.4 and a competitive cIoU of 50.4 on the full test set. When only one refinement point is used, the performance drops by 0.9 cIoU and 0.8 F1, indicating that a single point is insufficient to describe irregular socio-semantic boundaries. Using three points slightly improves cIoU to 50.6 but decreases F1 to 61.2, suggesting that additional points can expand region coverage but may also introduce less precise pixel-level decisions. In contrast, four-point refinement leads to a clear degradation, with 47.7 cIoU and 59.6 F1, which implies that redundant point prompts may disturb SAM2 when the explored evidence contains ambiguous map or satellite cues.

We further ablate the two reasoning stages. Removing the Enumeration stage causes the largest decline, reducing cIoU and F1 by 6.9 and 8.0 points, respectively. This result confirms that stable instance discovery and box anchoring provide the indispensable search space for subsequent boundary reasoning. Removing the Verification stage also decreases performance by 4.1 cIoU and 2.6 F1, showing that the initial candidates still require egocentric inspection, adjustment, and point-level correction before mask generation. 

\subsection{Case Study}
\begin{figure}[t!]
  \begin{center}
\includegraphics[width=.8\linewidth]{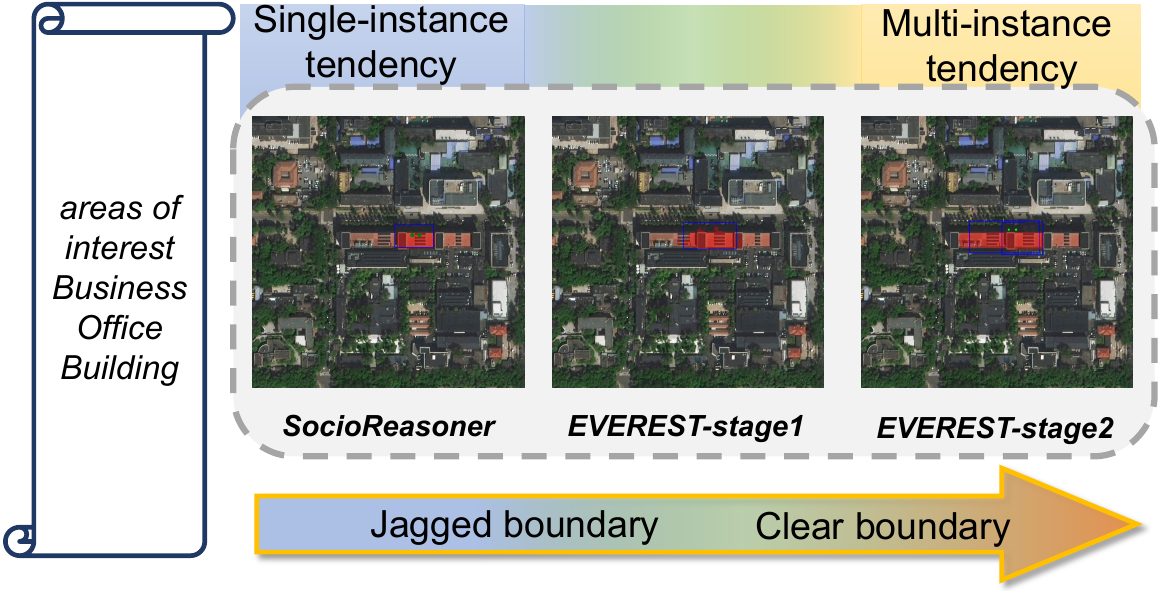}
  \end{center}
    \vspace{-20pt}
  \caption{Case study on SocioSeg}
  \vspace{-10pt}
  \label{fig:case-study}
  \vspace{-7pt}
\end{figure}

We conduct a case study on SocioSeg, with detailed qualitative comparisons presented in Fig. \ref{fig:case-study}. For Class-level entities such as Office building, SocioReasoner relies on reinforcement learning solely for domain adaptation and lacks an exploration mechanism, which leads to single-instance segmentation and ambiguous boundary delineation. \modelname produces more accurate and coherent segmentation results throughout the two-stage process.
\section{Conclusion}\label{sec:conclusion}
In this paper, we focus on the limitation of target-boundary misalignment in socio-entity segmentation and attribute this issue to the lack of active environmental exploration. To address this problem, we propose \modelname, which explores preselected bounding boxes from multiple perspectives and imposes pseudocode-based logical constraints to regularize the reasoning process. Experiments on real-world datasets demonstrate that the proposed method achieves state-of-the-art performance.

\clearpage
\bibliographystyle{IEEEbib}
\bibliography{strings}
\end{document}